\documentclass[aps,pre,twocolumn,
floats,amsmath,amssymb,showpacs,showkeys,preprintnumbers,amsmath,amssymb,floatfix]%
{revtex4}

\usepackage{graphicx}
\usepackage{dcolumn}
\usepackage{color}
\usepackage{subfigure}
\usepackage[dvipsnames]{xcolor}

\begin{document}

\title{Lacunarity Exponents}

\author{Michael Wilkinson$^{1,2}$, Marc Pradas$^{1,2}$, Greg Huber$^{3,2}$ and Alain Pumir$^{4,2}$}

\affiliation{\\ 
$^1$ School of Mathematics and Statistics,
The Open University, Walton Hall, Milton Keynes, MK7 6AA, England,\\
$^2$ Kavli Institute for Theoretical Physics, University of California, Santa Barbara, CA 93106, USA\\
$^3$ Chan-Zuckerberg Biohub, 499 Illinois St., San Francisco, CA 94158, USA,\\
$^4$ Laboratoire de Physique,
 Ecole Normale Sup\'erieure de Lyon, CNRS, Universit\'e de Lyon,
 F-69007, Lyon, France,
}

\begin{abstract}
Many physical processes result in very uneven, apparently random, distributions of 
matter, characterized by fluctuations of the local density over orders of 
magnitude. The density of matter in the sparsest regions can have a 
power-law distribution, with an exponent that we term the lacunarity exponent. 
We discuss a mechanism which explains the wide occurrence of these power laws, 
and give analytical expressions for the exponent in some simple models.
\end{abstract}

\pacs{05.40.-a,05.10.Gg,05.40.-a}
%05.45.-a         Nonlinear Dynamics and Chaos
% 05.10.Gg	Stochastic analysis methods (Fokker-Planck, Langevin, etc.)
% 05.40.-a	Fluctuation phenomena, random processes, noise, and Brownian motion 
\keywords{lacunarity, fractals, chaotic dynamics, Lyapunov exponent}

\maketitle

%\section{Introduction}
%\label{sec: 1}

Particulate matter can form highly inhomogeneous 
distributions, in which the density of particles can exhibit very 
large fluctuations, in many physical contexts.
Examples include the distributions of galaxies in the universe \cite{Jon+05}, 
the distributions of stars within galaxies \cite{Elm+01}, the distribution of debris 
floating on fluids \cite{Yu+91,Som+93} such as the surface of the ocean \cite{Coz+14}, the distribution 
of human populations \cite{Che11} and the distribution of small inertial particles, 
such as water droplets in clouds, in a turbulent flow \cite{Sha02}. Many of 
these distributions are fractal \cite{Man82,Fal90}, and can be characterised 
by a power law, with an exponent related to one or more fractal dimensions. 
In particular, the density correlation 
function has a power-law decay, which can be related to what is known
in dynamical systems theory as the correlation dimension, $D_2$ 
\cite{Gra+83,Ott02}.

\begin{figure}[t!]
\centering
\includegraphics[width=0.48\textwidth]{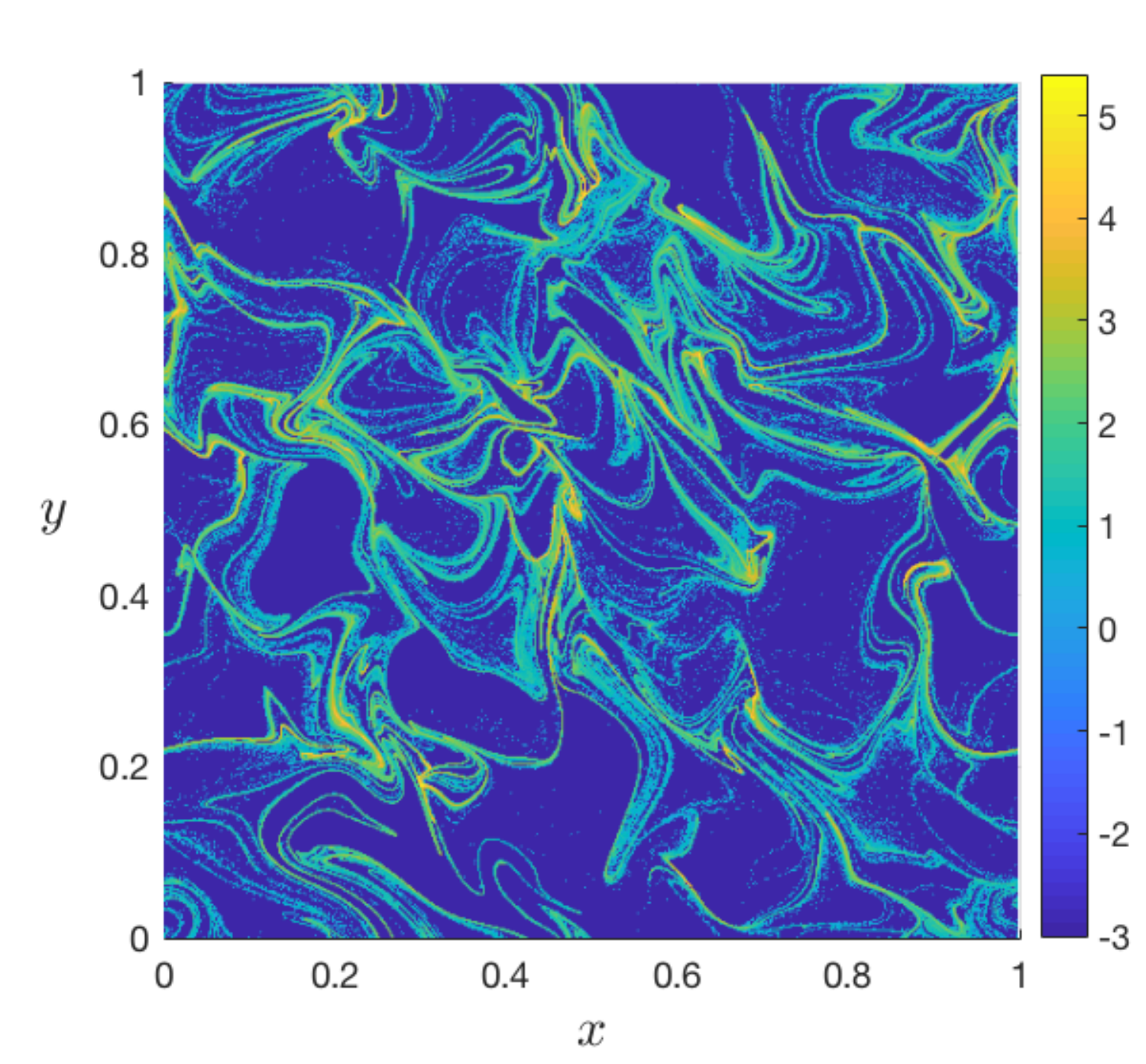}
\caption{(Color online.)
Distribution of particle positions in a two-dimensional random flow, mimicking 
the motion of debris on a sea surface. The density is inhomogeneous on length 
scales much larger than the correlation length $\xi$ of the equations of motion 
($\xi=0.05$ in this illustration).  The colour bar indicating density is in 
decimal log scale.}
\label{fig: 1}
\end{figure}

\begin{figure}%[h t b]
\centering
\includegraphics[width=0.46\textwidth]{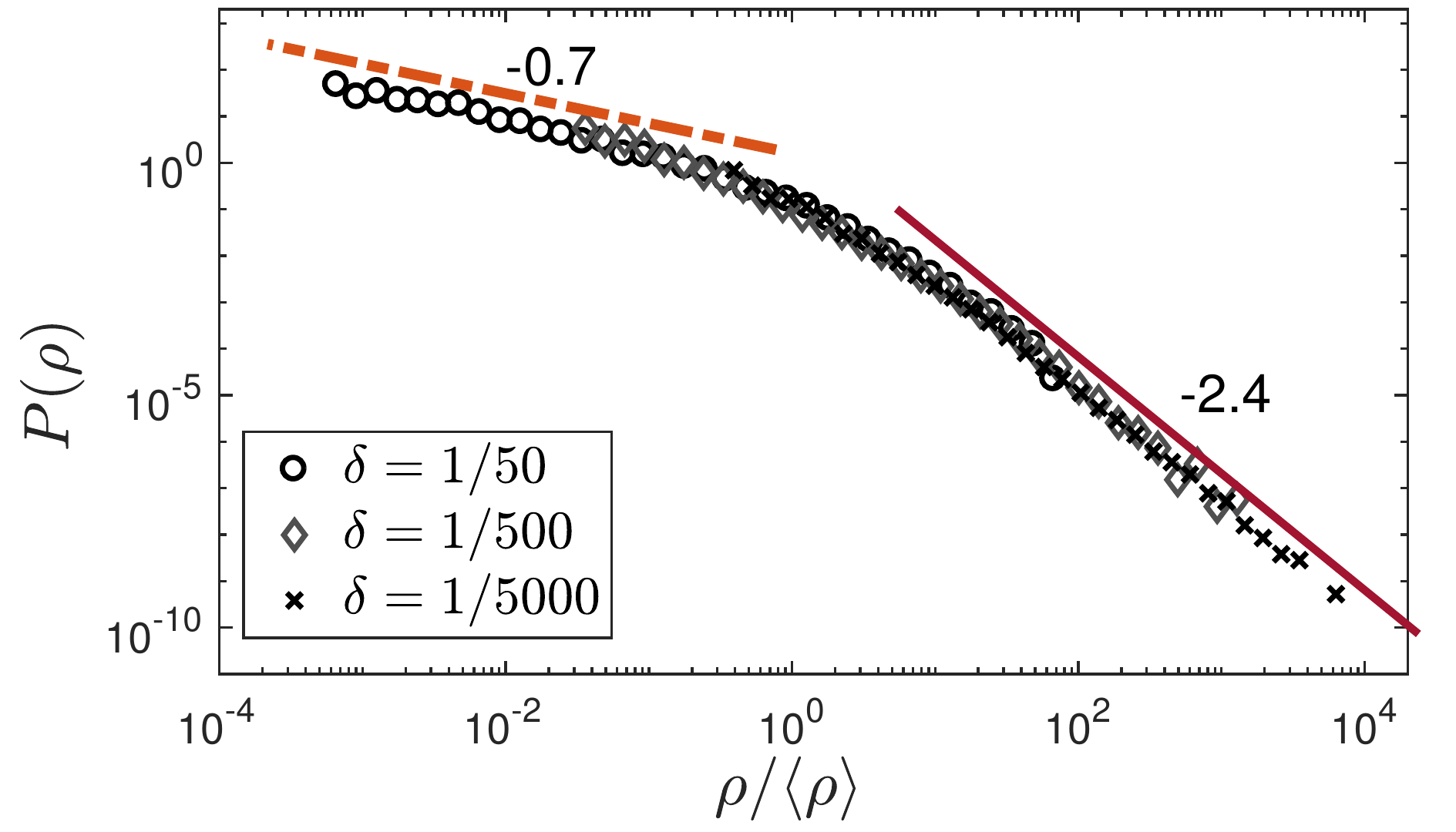}
\caption{ PDF of the particle density for the model illustrated in figure \ref{fig: 1}. 
There were $10^7$ particles, and the densities were evaluated using square regions of size $\delta$, 
much smaller than the correlation length $\xi=0.05$, and normalised by their mean value $\langle \rho\rangle$. The two asymptotes correspond to power laws with exponent $\alpha \approx 0.7$ (low densities) and $\beta\approx 2.4$ (high densities).}
\label{fig: 2}
\end{figure}

Fractal dimensions are characterisations of the densest regions of the distribution: the
definition of the dimension involves looking at the distribution of material inside a small
ball of radius $\varepsilon$, and considering a limit as $\varepsilon \to 0$ 
\cite{Man82,Fal90}. 
It can be equally valuable to understand the sparse regions 
of the distribution, but these have received relatively little 
attention in the literature. The prevalence of these low-density
regions is clearly illustrated by figure ~\ref{fig: 1}, which shows the 
distribution of particle position, determined from a two-dimensional,
compressible flow model of transport on the surface of the ocean 
(the equations of motion are discussed later: see equation (\ref{eq: 4.1}) below). 
There are large variations of particle density $\rho$, on length scales which are large compared to 
the correlation length $\xi$ of the model. The numerically determined probability density function
(PDF) of the particle density for this model, $P(\rho)$ is shown in figure \ref{fig: 2}. 
The plot of the distribution uses double-logarithmic scales, and there appear to be two linear asymptotes, indicating that the distribution approaches 
power-law forms, both at high and at low density, with two different 
exponents. We denote by $\alpha$ the exponent corresponding to low densities: 
\begin{equation}
\label{eq: 1.1}
P(\rho)\sim \rho^{-\alpha} ~~  {\rm as} ~~ \rho \to 0 
\end{equation}
The exponent that characterizes the high density regions is denoted  as $\beta$. 


The existence of a power-law at high 
densities is not very surprising, because it is consistent with the notion 
that chaotic dynamical systems can have fractal invariant measures
\cite{Ott02,Gra+83}. The occurrence of a power-law distribution at 
low densities, as documented in figure \ref{fig: 2}, is a distinct
phenomenon which deserves to be understood. 
In this paper we describe a general approach to understanding 
the sparse regions of the phase space of a broad class of complex dynamical systems. 
We show that there is a general mechanism explaining why the sparse regions can have a 
power-law  distribution of density, and discuss examples where the exponent $\alpha$
of the low-density distribution, which we term the \emph{lacunarity exponent}, can be estimated.
In support of our claim that the effect is widely observable, we note that 
evidence which supports equation (\ref{eq: 1.1}) has previously 
been presented in numerical and experimental studies of a few  
different systems with quite disparate  equations of motion 
(the examples that we know about, \cite{Bec+07,Lar+09,Pra+17, Kaw+17}, 
are discussed below).

The term \lq lacunarity' was introduced as a notion for characterising fractal sets by 
Mandelbrot \cite{Man82}, and various approaches to defining lacunarity have been 
explored \cite{Gef+83,Plo+96}, measuring the spatial inhomogeneity of a set (which need 
not be a fractal). Our definition of the lacunarity exponent characterises a very strong 
type of inhomogeneity, which we claim is a robust and widely observable phenomenon. 

\begin{figure}%[h t b]
\centering
\includegraphics[width=0.5\textwidth]{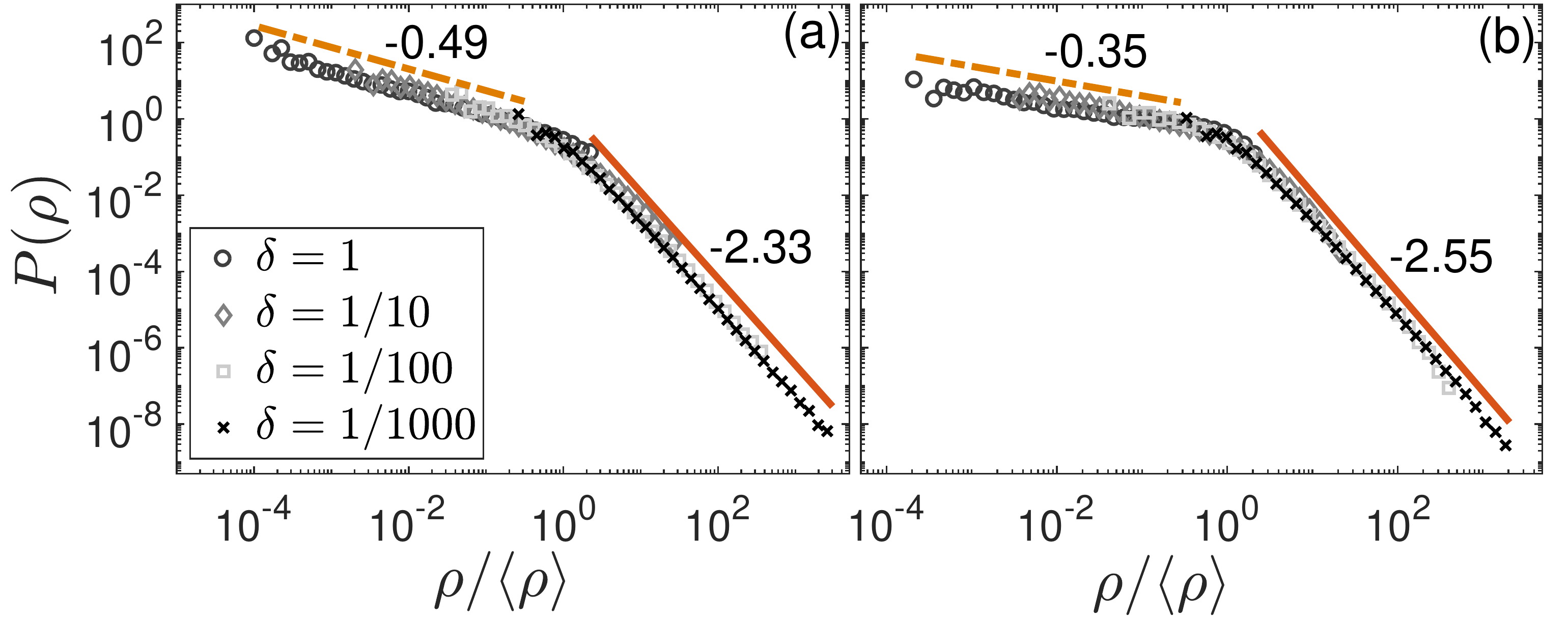}
\caption{
PDF of the density for the correlated random walk map, 
equations (\ref{eq: 1.1}) and (\ref{eq: 1.2}), for two different values of 
$\epsilon$, namely $\epsilon = 1.5\epsilon_c$ (a) and  
$\epsilon = 1.9\epsilon_c$ (b), which are expressed as multiples of the 
critical value $\epsilon_{\rm c}$ where 
the Lyapunov exponent vanishes. The PDF of the density exhibits a power-law, with 
different exponents, in both the low- and high-density limits. 
The dashed line and solid line correspond to the 
theoretical predictions given by Eq.~(\ref{eq: 2.4}) and Eq.~(\ref{eq: 2.8}), respectively.
}
\label{fig: 3}
\end{figure}

In order to explain the existence of the lacunarity exponent we consider the simplest
model for which we have observed its existence. This is the \emph{correlated 
random walk}, defined by
\begin{equation}
\label{eq: 1.2}
x_{n+1}=x_n+f_n(x_n)
\end{equation}
where the $f_n(x)$ are random, smooth functions, with $\langle f_n(x)\rangle=0$ and a correlation function of the form
\begin{equation}
\label{eq: 1.3}
\langle f_n(x)f_{n'}(x')\rangle
=\delta_{nn'}\epsilon^2\xi^2\exp\left(-\frac{(x-x')^2}{2\xi^2}\right)
\ ,
\end{equation}
where $\xi$ is the correlation length. 
This one-dimensional model is chaotic (that is, has a positive Lyapunov exponent) 
when $\epsilon$ exceeds a critical value $\epsilon_{\rm c}\approx 1.553$ \cite{Wil+03}. 
In figure \ref{fig: 3} we plot the PDF of the density of 
final points for this model after $N=100$ iterations, obtained by iterating a very 
dense distribution of initial points. The density $\rho$ in our numerical studies 
of this model is defined empirically, using the number of 
trajectories ${\cal N}$ in a randomly chosen 
interval of length $\delta$, by writing $\rho={\cal N}/ \delta$. 
The PDF of the density, $P(\rho)$, was determined by dividing the 
range of $x$ into bins of length $\delta$ (for the two-dimensional 
system illustrated in figures \ref{fig: 1} and \ref{fig: 2} 
we divided the coordinate space into 
squares of side $\delta$, and defined $\rho={\cal N}/\delta^2$).
The plot of the distribution 
shows that there are also two power-law asymptotes for this model, which are 
compared with theoretical predictions for the exponents $\alpha$ and $\beta$
which will be described below. We find that the exponents of both power laws
approach limits as the number of iterations $N$ increases and as the size of the
interval $\delta $ decreases, provided that the number of trajectories is sufficiently 
large (i.e. ${\cal N}\gg 1$). Figure \ref{fig: 3} contains data for four 
different values of $\delta $. We used a periodic realisation of the random functions $f_n(x)$, 
with period $L=5$, and with correlation length $\xi=0.5$. 
The total number of trajectories was $5\times10^4$, 
initially uniformly scattered over $[0,L]$. 
The exponents change continuously as the parameters of the model are varied. 

%\section{Asymptotes of the density distribution}
%\label{sec: 2}

Now we present arguments supporting the existence of these power-laws. 
For clarity, we start by considering the simplest case, 
which is a chaotic map in one dimension, such as that described 
by equations (\ref{eq: 1.2}) and (\ref{eq: 1.3}). If $\rho_N(x)$ is the density 
of particles after $N$ iterations, the density at $x$ after $N+1$ iterations is 
\begin{equation}
\label{eq: 2.1}
\rho_{N+1}(x)=\sum_{j=1}^K \left\vert\frac{\partial x_{N+1}}{\partial x_N}\right\vert^{-1}_{x_j} 
\rho_N(x_j)
\end{equation}
where the $x_j$ are the $K$ pre-images of $x$ at iteration $N$.

Because of the un-predictability 
of chaotic motion, we apply probabilistic methods. 
If there is only one pre-image, the value of $\rho$ is simply 
multiplied by $|\partial x'/\partial x|^{-1}$, which we treat as if it is a 
random factor. If there are multiple pre-images, 
the sum components are likely to be very different in magnitude, so that 
the density will be dominated by the largest term~\cite{PPW18}.  
Because of the multiplicative 
nature of the density mapping, the different branches can have dramatic 
differences in magnitude, so that approximating the sum by its largest 
term is a valid approximation when considering the limit of very high densities.
The statistical treatment of the map from $\rho_N$ to $\rho_{N+1}$ simplifies 
in both the low- and high-density limits, leading to a power-law for 
the probability density $P(\rho)$. We consider these in turn.

%\subsection{Low-density limit}
%\label{sec: 2.1}

We formulate our argument for maps which have Markovian properties. 
First we assume that the density at $x$ is very small. 
At the next iteration, where $x$ is mapped to $x'$, there 
may be other trajectories which reach $x'$. In general, the number $K$ of pre-images 
of $x'$ may be regarded as a random number, with a probability $P_K$ (note that 
$K$ is an odd number for continuous maps).

Consider how very small values of the density $\rho(x)$ are realised. Because the map 
is assumed to be chaotic, the values of factors by which the density is changed 
at each iteration, namely $|\partial x'/\partial x|^{-1}$, are expected to be typically 
less than unity, implying that the density will decrease upon iteration of the map.
This tendency to produce smaller densities is frustrated by the possibility of the 
map folding the pre-image space. In the limit as $\rho \to 0$, the occurrence 
of a typical folding event will increase the density to a typical value. 
Very small values of the density are, therefore, the result of trajectories 
which repeatedly escape folding. Because of the Markovian nature of a map 
such as equations (\ref{eq: 1.2}) and (\ref{eq: 1.3}), the probability of surviving for 
$m$ iterations without a folding event is $P_1^m$. In $m$ iterations without 
folding, the density changes by a factor $F_m$ which satisfies 
\begin{equation}
\label{eq: 2.2}
\langle \ln F_m\rangle =m
\bigg\langle \ln \biggl\vert\frac{\partial x'}{\partial x}\bigg\vert^{-1}\biggr \rangle_1
\end{equation}
where $\langle X\rangle_1$ is an average over trajectories which have only one pre-image.
The distribution of the logarithm of the density, $X=\ln \rho$, therefore satisfies
\begin{equation}
\label{eq: 2.3}
P(X)\sim \exp\left(\frac{\ln\,P_1}{\langle \ln |\partial x'/\partial x|^{-1}\rangle_1} X\right)
\ .
\end{equation}
This implies that there is a power-law distribution of $\rho$ of the form (\ref{eq: 1.1}) 
at small values, with lacunarity exponent
\begin{equation}
\label{eq: 2.4}
\alpha=1-\frac{\ln\,P_1}{\langle \ln |\partial x'/\partial x|^{-1}\rangle_1}
\ .
\end{equation}
In general it is extremely hard to make a theoretical calculation of either 
the numerator or the denominator in the expression for $\alpha$, equation (\ref{eq: 2.4}), 
because of the difficulty of incorporating the condition that there is no folding into the 
averaging procedure, but later we shall describe a simplified version of 
the model contained in (\ref{eq: 1.2}) and (\ref{eq: 1.3}) which does allow us to give an
exact expression for the exponent $\alpha$. The reasoning leading to the prediction of 
a power-law distribution for small values of $\rho$ is, however, very robust. We only used 
the assumption that there is a finite probability for a trajectory to have just one pre-image.

%\subsection{High-density limit}
%\label{sec: 2.2}

In the case where the density $\rho(x)$ greatly exceeds the typical value 
of the density, the contributions to $\rho(x')$ from other pre-images are almost 
certainly negligible, so that the transformed density is $\rho'(x')=\rho(x)/F$, where 
$F=|\partial x'/\partial x|$. The factor $F$ may be treated as a random variable 
with probability density $\pi(F)$.
After $N$ iterations the density $\rho$ varies extremely rapidly as a 
function of the coordinate $x$. Let $Q_N(\rho)$ be the probability that the density at a 
given point is less than $\rho$ after $N\gg 1$ iterations. At the next iteration, the cumulative 
probability depends upon the statistics of the 
sensitivity factor $F(x)=|\partial x_{N+1}/\partial x_N|$:
regions of density $\rho$ become regions of density $\rho'=\rho/F$, and intervals 
of length ${\rm d}x$ become intervals of length ${\rm d}x'=F{\rm d}x$. Because $\rho(x)$
fluctuates very rapidly as a function of $x$, a small region of length $\delta x$ has 
a range of different values of $\rho$, with cumulative weight $Q_N(\rho)\delta x$. At the next
iteration, the image of this interval has width $\delta x'=F\delta x$, with cumulative weight
$\delta x'Q_N(F\rho)$. Integrating over the distribution of $F$, the cumulative probability 
of $\rho$ is therefore iterated as follows:
\begin{equation}
\label{eq: 2.5}
Q_{N+1}(\rho)=\int_0^\infty {\rm d}F\ \pi(F)\, F\, Q_N(F\rho)
\ .
\end{equation}
Seeking a steady-state solution for the PDF of the density of the form 
$P(\rho)=\frac{{\rm d}Q}{{\rm d}\rho}\sim\rho^{-\beta}$ we find that
\begin{equation}
\label{eq: 2.7}
1=\int_0^\infty {\rm d}F\ \pi(F)\, F^{2-\beta}
\end{equation}
which is an implicit equation for the high-density exponent, $\beta$.

The existence of a power-law distribution of density in the high density limit 
can be related to the fractal properties of the attractor. 
One characterisation of the fractal properties is achieved by considering 
the expectation value $\langle {\cal N}(\varepsilon)\rangle$ of the number 
of trajectories within an interval of length $\varepsilon$ centred on a randomly 
chosen test trajectory:  $\langle {\cal N}\rangle \sim \varepsilon^{D_2}$, where
$D_2$ is termed the correlation dimension \cite{Gra+83}. In \cite{Wil+12}, it was shown 
that $D_2$ satisfies a relation closely related to (\ref{eq: 2.7}), implying that
\begin{equation}
\label{eq: 2.8}
\beta=D_2+2
\ .
\end{equation}

Our discussions have assumed that the empirical density, defined by writing 
$\rho={\cal N}/\delta$ for finite $\delta$, is compatible with the density 
defined mathematically by taking $\delta\to 0$. This is not guaranteed, because 
as the number of iterations of the map increases, the sensitivity of the final 
position $x_N$ to the initial coordinate increases exponentially with the number 
of iterations, $N$. This implies that, unless $N$ is small, the empirically determined 
density is given by an average of the value of $\rho(x)$ which fluctuates very rapidly 
over an interval. The mean number of trajectories in a randomly chosen interval 
of length $\varepsilon$ has the scaling $\langle {\cal N}\rangle\sim \varepsilon^{D_1}$, 
where $D_1$ is termed the information dimension. If the distribution 
of $P(\rho)$ has a 
finite mean value, we expect that the number of trajectories in an interval is proportional 
to the length of that interval, implying that $D_1=1$. The mean value is finite if $\beta>2$.
Equation (\ref{eq: 2.8}) implies that this condition is satisfied whenever $D_2>0$. This holds
whenever the motion is chaotic (that is, the map has positive Lyapunov exponent \cite{Wil+12}). 

%\section{An exactly solvable model}
%\label{sec: 3}

In the general case the equation for the exponent $\alpha$, (\ref{eq: 2.4}), is not susceptible 
to further analysis and must be solved numerically. We describe one special case of the map 
(\ref{eq: 1.2}) where an explicit expression for $\alpha$ has been obtained. Consider the map 
\begin{equation}
\label{eq: 3.1}
x_{n+1}=x_n+F(x-\phi_n)
\end{equation}
where $F(x)=F(x+1)$ is a periodic function, and where $\phi_n$ is a random 
number which is uniformly distributed on $[0,1]$. We consider the case where 
$F(x)$ is piecewise linear:
\begin{equation}
\label{eq: 3.2}
F(x)=
\left\{ 
\begin{array}{cc}
gx & 0\le x\le \frac{1}{2} \cr
g(1-x) & \frac{1}{2}\le x \le 1
\end{array}
\right.
\end{equation}
(where $g>0$), so that the function $F$ is continuous and piecewise linear, with 
gradients $\pm g$. For this model, the forward map (\ref{eq: 3.1}) has gradients 
with magnitude $1+g$ or $1-g$, both with with probability $\frac{1}{2}$. 
The Lyapunov exponent of the model is
\begin{equation}
\label{eq: 3.3}
\lambda=\frac{1}{2}\left[\ln |g+1|+\ln |g-1|\right]
\end{equation}
so that the model is chaotic ($\lambda >0$) when $g>\sqrt{2}$. 

Now consider how to compute the exponents $\alpha$ and $\beta$ which 
characterise the low- and high-density limits of the PDF of $\rho$. 
When $g>1$, a point may have multiple pre-images. When 
$g<3$, any point has either a single pre-images, with probability $P_1$, or else 
three pre-images, with probability $P_3=\frac{(g-1)}{2}$. 
For this model, points which have just one pre-image all have 
the same expansion factor, namely $g+1$, so that the low-density 
exponent is
\begin{equation}
\label{eq: 3.4}
\alpha=1+\frac{\ln (1-P_3)}{\ln (g+1)}=1+\frac{\ln[(3-g)/2]}{\ln(g+1)}
\ .
\end{equation}
Equation (\ref{eq: 2.7}) determining the high-density exponent, $\beta=D_2+2$, becomes
\begin{equation}
\label{eq: 3.5}
(g+1)^{(2-\beta)}+(1-g)^{(2-\beta)}=2,
\end{equation}
which is an implicit equation that must be solved numerically.
Comparing equations (\ref{eq: 3.4}) and (\ref{eq: 3.5}), it is clear that there 
is no simple relationship between the exponents $\alpha$ and $\beta$.

\begin{figure}%[h t b]
\centering
\includegraphics[width=0.49\textwidth]{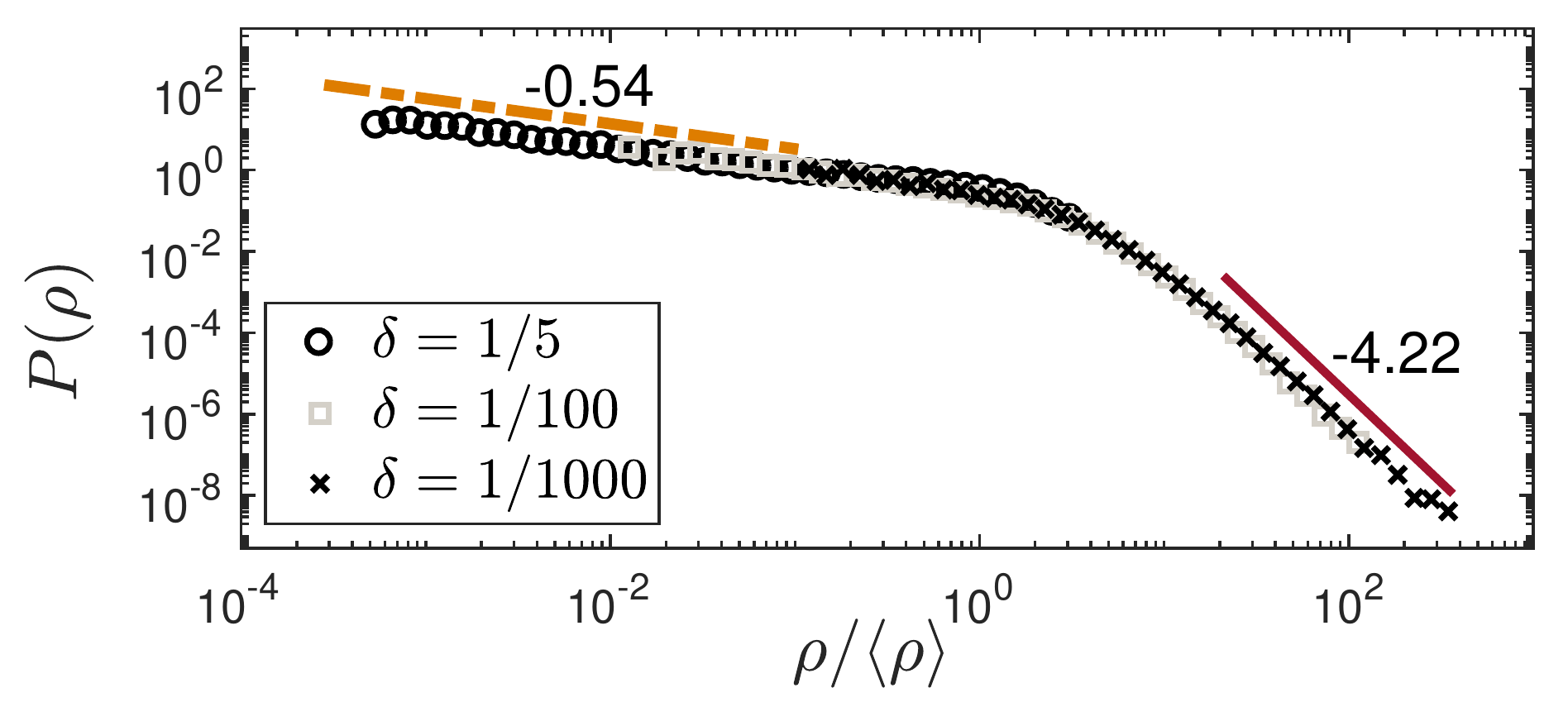}
\caption{
PDF of the density for the piecewise linear correlated random walk map, 
Eqs.~(\ref{eq: 3.1}) and (\ref{eq: 3.2}), with $g=\frac{7}{4}$, where (\ref{eq: 3.4})
and (\ref{eq: 3.5}) give $\alpha=0.54$ (dashed line) and $\beta=4.22$ (solid line), respectively.}
\label{fig: 4}
\end{figure}

We tested our predictions numerically. Figure \ref{fig: 3} shows a comparison 
between the numerically computed density PDF and the values of the exponents 
obtained from (\ref{eq: 2.4}) and (\ref{eq: 2.8}). For $\epsilon/\epsilon_{\rm c}=1.5$, 
these equations predict $\alpha=0.49$ and $\beta=2.33$ respectively, whereas 
for $\epsilon/\epsilon_{\rm c}=1.9$, they predict $\alpha=0.35$ 
and $\beta=2.55$, in very good agreement 
with the data in figure \ref{fig: 3}. Figure \ref{fig: 4} shows the numerically 
determined PDF of the piecewise linear model, comparing the exponents 
with (\ref{eq: 3.4}) and the solution of (\ref{eq: 3.5}). We can see there is clear 
evidence for a power-law at both high and low densities, confirming the existence of the lacunarity 
exponent, in agreement with the theory.

%\section{Generalisations}
%\label{sec: 4}

We mentioned in the introduction that power-law distributions of small 
densities had been observed in a variety of other contexts. 
Bec {\sl et al.} \cite{Bec+07} showed that the density of inertial particles in simulations 
of three-dimensional turbulence has a lower-law at low densities.
Larkin {\sl et al.} \cite{Lar+09} have exhibited a density distribution for particles 
advected on the surface of a turbulent flow in a water tank. The density distribution 
shows two power-laws, at high and low densities.
Kawagoe {\sl et al.} \cite{Kaw+17} discuss the distribution of weights for a model of 
random \lq trails'. The distribution is asymptotic to a power-law in the limit 
where the weight of a trail approaches zero, and the effect is quite distinct from the growth of 
voids observed close to the directed percolation transition \cite{Hub+95}.
Finally, the present authors previously reported \cite{Pra+17} data quite closely 
related to figure \ref{fig: 3}, showing high- and low-density power laws for a 
one-dimensional model of inertial particles.
These observations concern models for the distribution of particles with quite 
different equations of motion, in different numbers of space dimensions. 
With the exception of \cite{Kaw+17}, which uses an exact solution for a specific model, 
these papers do not offer a clear explanation for the power-law density distribution.
This begs the question as to whether there is a common mechanism 
which implies power-law behaviour. 

The argument that we presented for the existence of the lacunarity exponent 
in the one-dimensional case is readily extended to a wide range of chaotic 
and random dynamical systems. There are two essential elements to the argument, 
embodied in equations (\ref{eq: 2.2}) and (\ref{eq: 2.3}) respectively, which 
can be generalised. Firstly, there must be a finite fraction of trajectories for 
which the volume of the pre-image of a small neighbourhood contracts 
exponentially as we go backwards in time. Secondly, there must be a mechanism 
which ensures that the pre-images do not continue to contract indefinitely,
so that the sequence of contractions is broken. If the probability that this 
mechanism does not occur decreases exponentially as we go backwards in
time, then the argument supporting equation (\ref{eq: 2.4}) implies 
that $P(\rho)$ has a power-law distribution as $\rho\to 0$. 
In the case of the one-dimensional map, the mechanism for breaking the series of 
pre-image contractions is the existence of multiple pre-images.

These arguments apply directly to most of the previously reported examples, 
\cite{Bec+07,Lar+09,Kaw+17,Pra+17}. 
The exponential reduction of density for a single trajectory as a 
function of the number of iterations is a feature of  all these systems: in most cases it 
is a result of the exponential separation of nearby trajectories, but in \cite{Kaw+17} 
it is the result of trajectories randomly splitting, reducing the weight in each daughter 
trajectory.
In all of the examples in \cite{Bec+07,Lar+09,Kaw+17,Pra+17} except \cite{Lar+09}, 
the different trajectories can have multiple pre-images, and the probability of 
avoiding trajectories crossing or combining may be assumed to decrease 
exponentially as a function of time. 
The system considered by Larkin {\sl et al.} \cite{Lar+09} involving particles floating 
on the surface of a liquid which is undergoing a turbulent or complex flow, 
is different: the equation of motion is $\dot {\mbox{\boldmath$x$}} =\mbox{\boldmath$u$} (\mbox{\boldmath$x$},t)$, 
with a random velocity field $\mbox{\boldmath$u$} (\mbox{\boldmath$x$},t)$. 
This system has a unique time-reversed motion. In this case a different 
mechanism for breaking the contraction of the pre-images may play a role. 
The pre-images of a small disc can expand in one direction, while 
their area contracts. The 
exponential contraction of the pre-image area breaks down when 
stretching makes its extent too large, so the linear 
approximation becomes inappropriate. 
In our numerical 
illustration of a two-dimensional flow, figures \ref{fig: 1} and \ref{fig: 2}, 
we used a simplified model
\begin{equation}
\label{eq: 4.1}
{\mbox{\boldmath$x$}}(t+\Delta t) ={\mbox{\boldmath$x$}}(t)+\mbox{\boldmath$u$} (\mbox{\boldmath$x$},t)
\Delta t
\ .
\end{equation}
The velocity field was constructed by writing $\mbox{\boldmath$u$}
=\mbox{\boldmath$\nabla$}\wedge \mbox{\boldmath$\psi$} +\eta \mbox{\boldmath$\nabla$}\phi $
where $\phi$ and $\psi$ have the same isotropic, homogeneous statistics and are independent
of each other, and where $\eta$ is a parameter which controls the compressibility. 
In the simulations we used $\eta=0.5$ and a large value of $\Delta t$
($\Delta t=0.5$) to ensure that the forward and 
time-reversed flows have different statistical properties. 

%\section{Conclusions}
%\label{sec: 5}

In summary, we have argued that dynamical processes can produce regions of extremely 
sparse trajectories, characterised by a power-law distribution of density, parametrised by 
the lacunarity exponent, $\alpha$. This property of having a power-law distribution of small 
densities is complementary to the power-law distribution of high-density regions, 
which is associated with a fractal dimension. 
We have described a robust mechanism explaining the existence of this power law, 
and shown how the exponent can be computed for one-dimensional systems. 
The existence of a lacunarity exponent is consistent with existing observations 
on a variety of dynamical systems, and it may prove to be a widely observable 
and quantifiable property.

{\sl Acknowledgements}. The authors are grateful to the
Kavli Institute for Theoretical Physics for support, where
this research was  supported in part by the National Science Foundation
under Grant No. PHY11-25915.

Author email addresses:

\indent \ \ \ marc.pradas@open.ac.uk

\indent \ \ \ alain.pumir@ens-lyon.fr

\indent \ \ \ greg.huber@czbiohub.org

\indent \ \ \ m.wilkinson@open.ac.uk

\end{document}